\documentclass[twocolumn,preprintnumbers,prb,fleqn]{revtex4}
\usepackage{graphicx,amsfonts,amsbsy}
\usepackage[dvips]{epsfig,color}
\begin{document} 
%\title{$T_c$ reduction due to spin clustering in diluted magnets}
%\title{Competition between cluster and bulk couplings in diluted magnets}
\title{Clustering induced suppression of ferromagnetism in diluted magnets}
\author{Avinash Singh,$^{1,2,*}$}
%\affiliation{$^1$Institut f\"{u}r Physik, Humboldt-Universit\"{a}t zu Berlin, Newton str. 15, D-12489 Berlin}
\affiliation{$^1$Max-Planck-Institut f\"{u}r Physik Komplexer Systeme, N\"{o}thnitzer str. 38, D-01187 Dresden}
\affiliation{$^2$Department of Physics, Indian Institute of Technology Kanpur - 208016}
\email{avinas@iitk.ac.in}
\begin{abstract}
Ferromagnetism in diluted magnets in the compensated regime $p \ll x$ 
is shown to be suppressed by the formation of impurity spin clusters.
The majority bulk spin couplings are shown to be considerably weakened
by the preferential accumulation of holes in spin clusters,
resulting in low-energy magnon softening and enhanced low-temperature decay of magnetic order.
A locally self-consistent magnon renormalization analysis of spin dynamics shows that  
although strong intra-cluster correlations tend to prolong global order,  
$T_c$ is still reduced compared to the ordered case.
\end{abstract}
\maketitle
%\section{Introduction}
Diluted magnetic semiconductors (DMS) such as Ga$_{1-x}$Mn$_x$As exhibit a novel carrier-induced ferromagnetism
where doped mobile carriers mediate magnetic interactions between the $S=5/2$ Mn$^{++}$ 
impurity spins,\cite{matsu,dietl,ohno3,timm,dassarma}
with magnetization and Curie temperature $T_c$ increasing 
with carrier concentration.\cite{boukari,ku,edmonds,nazmul,liu}
While as-grown samples of Ga$_{1-x}$Mn$_x$As with high Mn concentration exhibit
a large degree of compensation, where the hole density $p$ is a small fraction 
of the Mn impurity concentration $x$, recent progress in fabrication and annealing
has yielded nearly compensation free samples, especially at low Mn concentrations.

Strong thermal decay of magnetization observed in these diluted magnets has been characterized recently 
in terms of spin-wave properties.\cite{sperl} 
Magnetization studies of Ga$_{1-x}$Mn$_x$As samples with different Mn content 
(thickness about 50nm, Mn content ranging from $2\%$ to $6\%$)
using SQUID (superconducting quantum interference device) measurements, 
with linear extrapolation from a 0.3-0.4 T magnetic field to overcome 
anisotropy fields which result in spin-reorientation transitions,\cite{sawicki} 
have found the spontaneous magnetization to be well described by the Bloch form $M(T)=M_0(1-BT^{3/2})$, 
with a spin-wave parameter $B \sim 1-3\times10^{-3}\; {\rm K}^{-3/2}$ 
which is about two orders of magnitude higher than for Fe and FeCo films.\cite{sperl,kipferl} 
The calculated values of spin-wave stiffness constant $D$ were found to be of 
same order of magnitude as obtained from magnetic Kerr measurements using pump-probe setup
of standing spin waves in $\rm Ga_{1-x}Mn_x As$ thin films.\cite{notredame} 

Recent theoretical studies of magnon properties 
in the diluted ferromagnetic Kondo lattice model (FKLM)
provide quantitative understanding of this large magnitude of $B$ in terms of 
a strong enhancement in density of low-energy magnetic excitations at higher dilution.\cite{disklm}
The resulting strong thermal decay of magnetization 
was found to be in good agreement with the Bloch form at low temperature, 
with $B$ of same order of magnitude as obtained in the squid magnetization measurements.
%Essentially, a constant magnon density of states with a low-energy cutoff yields a 
%$T\ln T$ type decay of magnetization, which is functionally similar to the $T^{3/2}$ form. 

%What is the origin of enhanced density of low-energy magnons in the diluted FKLM?
Generally, presence of competing antiferromagnetic (AF) interactions is a source of 
magnon softening, due to the negative-energy contributions from AF bonds 
when spins are twisted away from the collinear ferromagnetic arrangement. 
Indeed, close to the limit $p \sim x$ of half filling, 
competing AF interactions become strong enough
to destabilize the ferromagnetic state in the diluted FKLM.
In the compensated regime $p \ll x$, however, the relevant mechanism is quite different.

In this paper we highlight a different aspect of competing interactions in diluted magnets --- 
a competition between cluster and bulk spin couplings. 
In the compensated regime $p \ll x$, 
the formation of spin clusters (due to impurity positional disorder) 
and accumulation of doped holes in these impurity-rich regions 
significantly deprives the majority bulk spins of holes.
Because of the crucial compensation condition  $p \ll x$, 
even if cluster spins constitute a small fraction of the total,
they can accumulate a majority of doped holes.
While the cluster couplings are therefore dramatically enhanced, 
yielding high-energy cluster-localized magnon modes and strong intra-cluster ordering, 
the couplings between bulk spins, which are essentially responsible
for long-range ferromagnetic order, are significantly weakened,
resulting in low-energy magnon softening and enhanced thermal decay of magnetization.
These effects of impurity clustering on spin couplings and dynamics have been 
qualitatively discussed earlier within the impurity-band representation,
applicable near the metal-insulator transition.\cite{dis2}

Our objective here is to quantitatively connect, within a more general and unified framework, 
the different relevant aspects  
--- disorder, clustering, fermion polarizations, spin couplings, magnons, and spin dynamics.
With respect to magnetic effects of impurity clustering in DMS, 
as distinguished from comparison with the ordered case, earlier studies have found
softening of low-energy magnon modes, weakening (strengthening) of bulk (cluster) couplings, 
and formation of high-energy cluster-localized modes,\cite{dis2,dms,squid} 
formation of ferromagnetic droplets but no significant effects on $T_c$ in Monte Carlo calculations,\cite{mayr,prior} and reduction of $T_c$ in density functional\cite{sand} 
and local spin-density approximation\cite{xu} studies 
involving ab-initio evaluation of exchange parameters and subsequent statistical analysis.

The interplay between itinerant carriers in a partially filled band 
and localized impurity moments is conventionally studied within the diluted FKLM
\begin{equation}
H = t \sum_{i,\delta,\sigma} a_{i,\sigma}^\dagger a_{i+\delta,\sigma} 
- \frac{J}{2} \sum_I{\bf S}_I.{\mbox{\boldmath $\sigma$}}_I
\end{equation}
involving itinerant electrons hopping on host sites $i$
and a local exchange coupling between the localized impurity spin ${\bf S}_I$ 
and itinerant electron spin ${\mbox{\boldmath $\sigma$}}_I$.
Recently, magnon properties in the mixed spin-fermion model (1) were discussed extensively
with respect to variations in dilution $x$, exchange coupling $J$, impurity energy $\epsilon_d$, 
and hole concentration $p$ 
within an exact treatment of impurity positional disorder
amd a non-perturbative treatment of exchange coupling.\cite{disklm}
The magnon propagator was studied in the random phase approximation (RPA), 
where the bubble diagrams, representing repeated interactions between impurity spins
mediated by the particle-hole bubble, 
provide the lowest-level spin-rotationally-symmetric treatment of transverse spin fluctuations
where the Goldstone mode is explicitly preserved. 

\begin{figure}
\hspace*{-3mm}
\includegraphics[width=100mm]{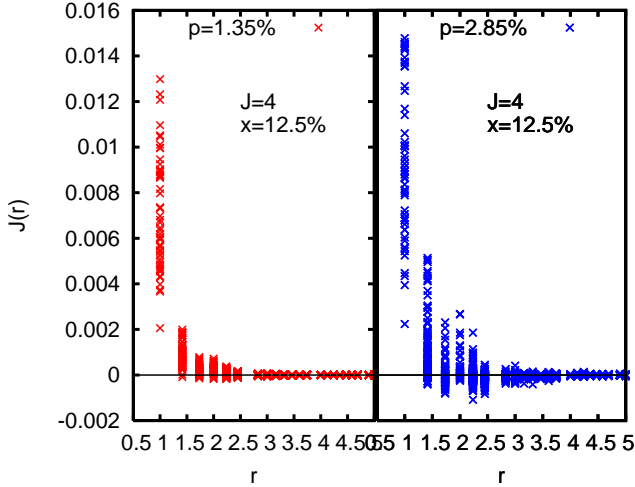}
\caption{(Color online) Distribution of spin couplings (single configuration) at two doping concentrations,
showing large spread in the strongly enhanced cluster couplings  
and distinctly non-RKKY nature in the compensated regime $(p\ll x)$.}
\end{figure}

Due to spin-rotation symmetry of (1),
the impurity spin dynamics is naturally described by an isotropic interaction
$-\sum_{IJ} J_{IJ} {\bf S}_I . {\bf S}_J $ between impurity spins,
and at the RPA level, the (weakly-dynamical) carrier-mediated spin couplings are given by 
\begin{equation}
J_{IJ}(\omega)= \frac{J^2}{4}[\chi^0(\omega)]_{IJ}
\end{equation}
in terms of the particle-hole bubble
\begin{eqnarray}
[\chi^0(\omega)]_{IJ} &=& 
i \int  \frac{d\omega'}{2\pi}[G^{\uparrow}(\omega')]_{IJ}[G^{\downarrow}(\omega'-\omega)]_{JI} \\ 
&=& \sum_{l,m}
\frac{\psi_{l\uparrow}^I \psi_{l\uparrow}^J \psi_{m\downarrow}^I \psi_{m\downarrow}^J }
{E_{m\downarrow} - E_{l\uparrow} + \omega} \; f_{l\uparrow}(1-f_{m\downarrow}) \nonumber \\
&+&
\sum_{l,m}
\frac{\psi_{l\uparrow}^I \psi_{l\uparrow}^J \psi_{m\downarrow}^I \psi_{m\downarrow}^J }
{E_{l\uparrow} - E_{m\downarrow} - \omega} \; (1-f_{l\uparrow})f_{m\downarrow}  
\end{eqnarray}
evaluated by integrating out the fermions in the broken-symmetry state.
%with Fermi functions $f_{l\uparrow}$ and $f_{m\downarrow}$.
Spin-fluctuation corrections to spin couplings can be incorporated 
by including self-energy and vertex corrections in the particle-hole bubble
within a spin-rotationally-symmetric scheme,\cite{vertex} 
but are suppressed by the factor $1/S$. 
The above isotropic spin interaction neglects magnetic anisotropy arising from spin-orbit 
interaction in the host semiconductor,\cite{schliemann,zarand,brey,berciu04} 
and the recently observed strain-induced uniaxial anisotropy 
which is dependent on hole concentration and temperature.\cite{sawicki}

For simplicity, we consider (1) on a simple-cubic lattice, with periodic boundary conditions. 
The $N_m$ magnetic impurities are placed randomly on a fraction ($I$) of the $N=L^3$ host sites ($i$),
with impurity concentration $x=N_m/N$. 
We consider positive nearest-neighbour hopping $t$, and set $t=1$ as the unit of energy scale. 
For the fully polarized, collinear ferromagnetic ground state at $T=0$, 
stability of which is confirmed from absence of negative-energy magnon modes,
the fermion eigenvalues $\{E_{l\sigma}\}$ and wave functions $\{\psi_{l\sigma}\}$ 
are obtained by exact diagonalization of the $N\times N$ fermion
Hamiltonian with effective impurity potentials $\mp JS/2$ for the two fermion spins.
Furthermore, we consider the saturated ferromagnetic state with a fully occupied spin-$\uparrow$ band and hole doping only in the pushed-up spin-$\downarrow$ band
($p\equiv N_\downarrow^{\rm holes}/2N$),
so that only the first term in Eq. (4) contributes to the particle-hole bubble.
For $p \sim x$ or lower values of $J,x$, we do obtain negative-energy magnon modes, 
indicating noncollinear ferromagnetic ordering, as found in earlier studies.\cite{macdonald}
As magnon energies are very low compared to the Stoner gap,
the spin couplings are only weakly dynamical,
and hence we have set $\omega=0$ in Eq. (2).
Although their nearly static nature is similar to that of RKKY interaction, 
the effective spin couplings are strongly non-perturbative in character 
due to the $J,x$ dependence of the wavefunctions and eigenvalues in Eqs. (3) and (4). 

\begin{figure}
\hspace*{-3mm}
\includegraphics[width=90mm]{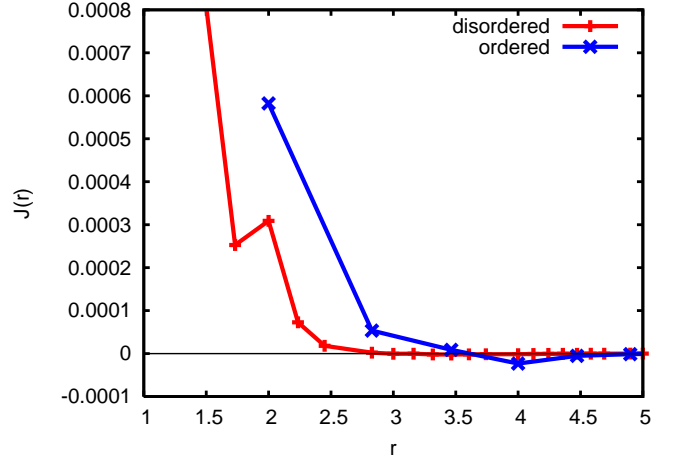}
\caption{(Color online) Substantially weakened bulk spin couplings at separations 
$\gtrsim 2$ compared to the ordered case,
for a $N=10^3$ system with $J=4$, $x=1/8$, and $p=1.35\%$,
averaged over 20 configurations.}
\end{figure}

Figure 1 shows the distribution of impurity spin couplings $J_{IJ}$ obtained from (2,4) 
as a function of impurity separation $r\equiv |{\bf r}_I - {\bf r}_J|$,
for two different hole concentrations $p/x \approx 1/10$ and $\approx 1/4$. 
Relatively very strong spin couplings are generated between cluster spins 
having small separations 1 and $\sqrt{2}$, 
which account for the high-energy cluster-localized magnon modes.\cite{disklm}
A most striking feature is that for the same impurity separation, 
the coupling magnitudes are spread over a large range.
The coupling between two impurity spins is thus not merely a function of their separation,
but actually depends on the whole configuration,
suggesting shades of a complex system.\cite{complex}
Indeed, for same $x,p,r$, the bulk spin couplings in the disordered case are 
strongly weakened compared to the ordered case. 
This feature is absent in certain spin-only models where the couplings are fixed functions of the separation,
quite independent of disorder.

Furthermore, while AF couplings are distinctly present for $p/x \approx 1/4$
and indeed the collinear ferromagnetic state is unstable,
in the compensated regime $p/x \approx 1/10$, where the ferromagnetic state is quite stable,
the couplings are seen to be dominantly ferromagnetic.
The plot of configuration-averaged couplings (Fig. 2) confirms 
that the couplings remain ferromagnetic and do not change sign. 
This distinctly non-RKKY behaviour of spin couplings in diluted magnets
has been highlighted recently in a detailed study.\cite{bouzerar}

\begin{figure}
\hspace*{-3mm}
\includegraphics[width=90mm]{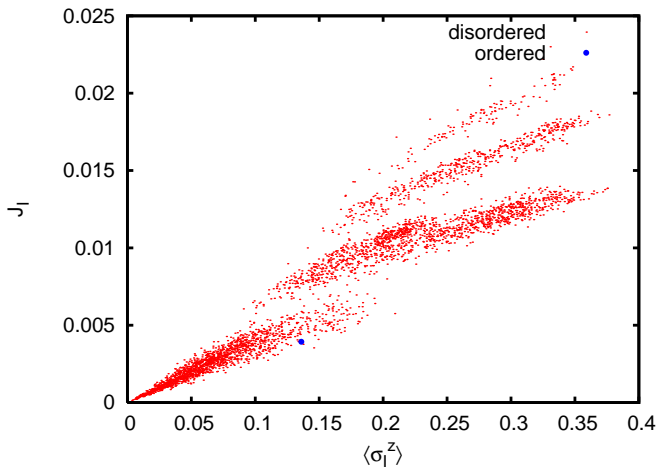}
\caption{(Color online) Correlation between impurity-site fermion polarization $\langle \sigma_I ^z\rangle$
and the local measure $J_I \equiv \sum_{J\ne I} J_{IJ}$ of spin couplings,
obtained for 20 configurations of a $N=10^3$ system with $J=4$, $x=1/8$, and $p=1.35\%$. 
Relative to the ordered case (bold dot), 
significantly enhanced (reduced) fermion-spin densities and impurity-spin couplings 
correspond to cluster (bulk) spins.}
\end{figure}

In order to examine the effect of impurity clustering on bulk spin couplings
we make a comparison with an ordered diluted case. 
Fig. 2 shows the spin couplings for disordered (configuration-averaged over 20 configurations)
and ordered cases for the same impurity concentration and hole doping. 
Here the ordered case corresponds to a superlattice arrangement of impurities 
with impurity spacing 2 and concentration $x=1/8$. 
The bulk spin couplings involving average separations $\gtrsim 2$ are clearly
considerably weakened, which is directly responsible 
for the observed low-energy magnon softening with dilution.\cite{disklm}

\begin{figure}
\hspace*{-3mm}
\includegraphics[width=90mm]{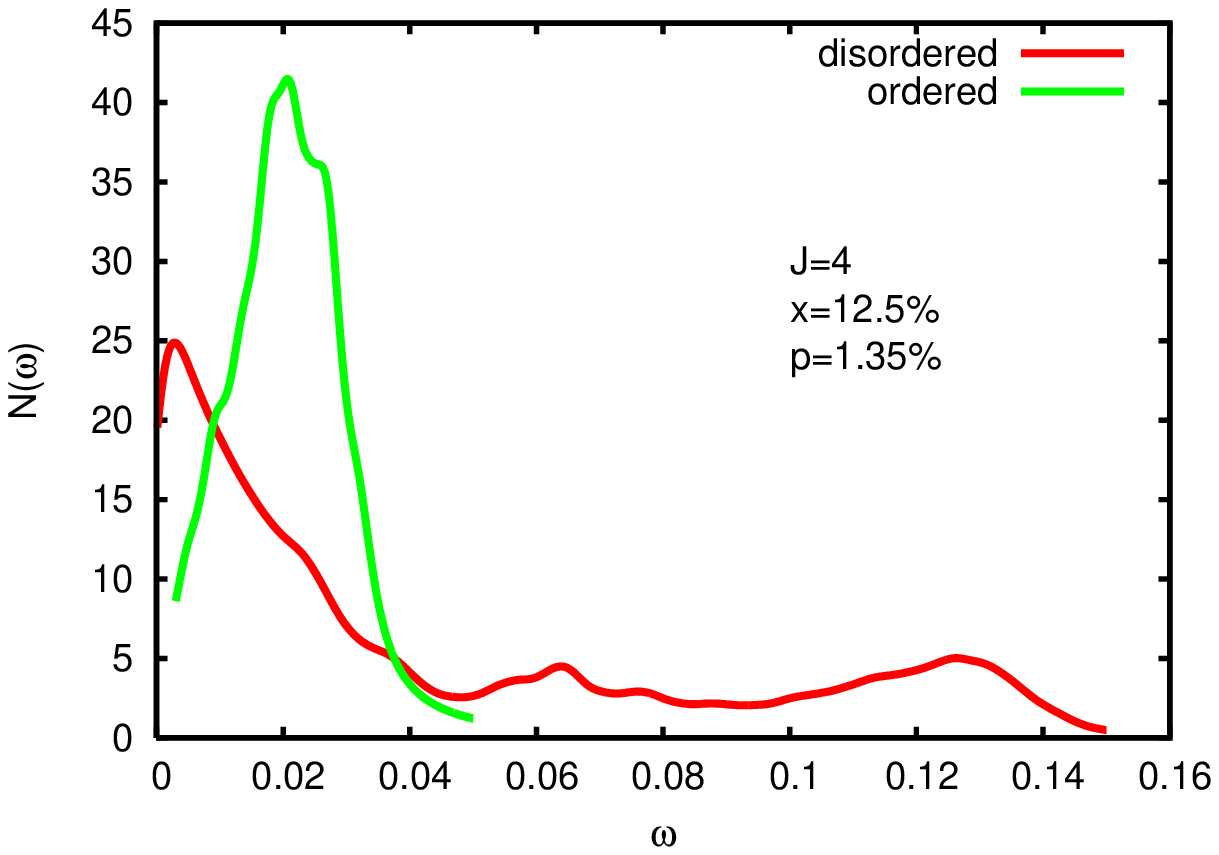}
\hspace*{-3mm}
\includegraphics[width=90mm]{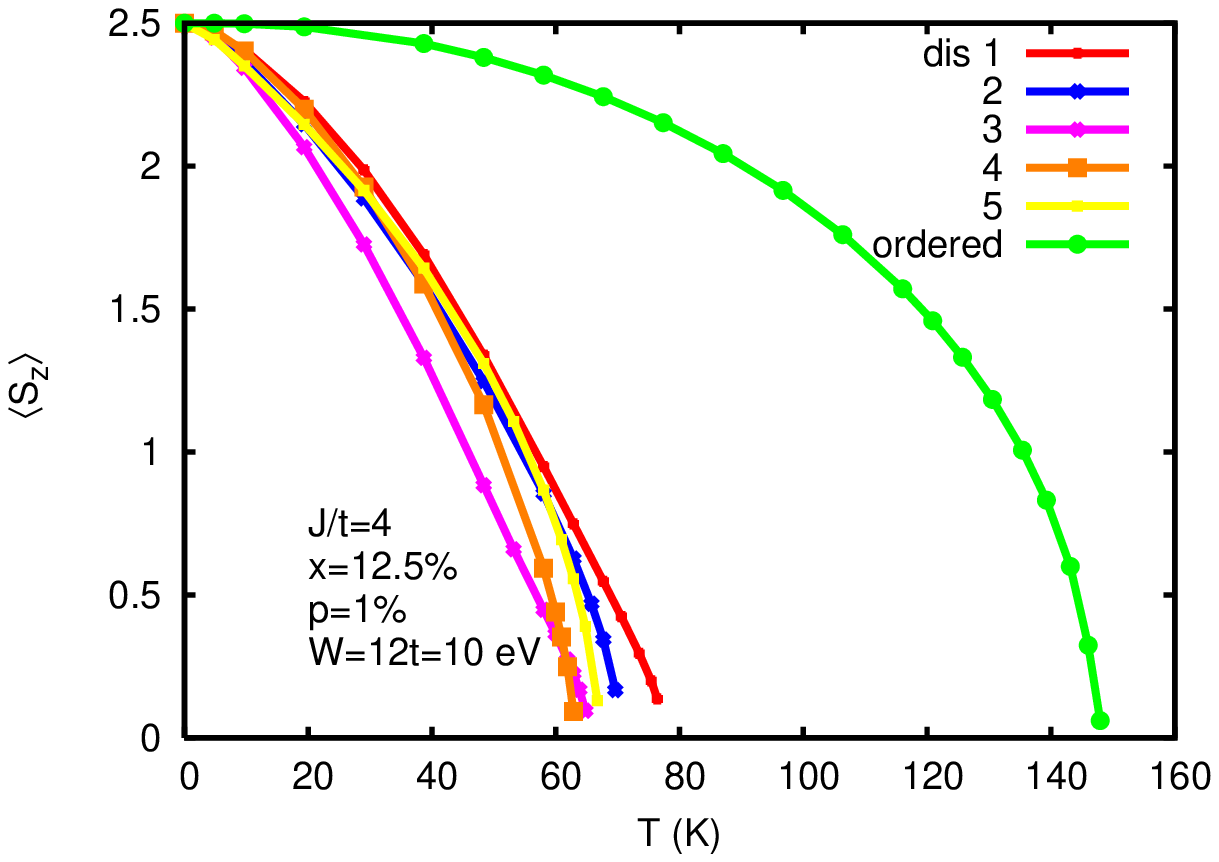}
\caption{(Color online) The strong low-$T$ decay of magnetization, 
with a subsequent marginally paramagnetic behaviour 
correspond to the enhanced low-energy magnon DOS and 
tendency of intra-cluster correlations to prolong order,
shown for 5 disorder configurations of a $N=12^3$ system 
in comparison with the ordered case at exactly same dilution $x=1/8$.}
\end{figure}

Turning now to include fermions, and the main result of this paper, 
Figure 3 highlights the essential character of carrier-induced spin couplings 
in the presence of disorder in terms of a direct correlation between 
local fermion polarization $\langle \sigma_I ^z\rangle$ and the local measure 
$J_I \equiv \sum_{J\ne I} J_{IJ}$ of spins couplings.
Fermion polarization values larger than the ordered-case value 
correspond to cluster sites which attract a disproportionately higher concentration of holes,
yielding proportionately larger spin couplings. 
The remaining bulk sites are thus left with a hole deficit,
resulting in proportionately weaker spin couplings. 
The compensation condition $p \ll x$ plays a crucial role here 
because even if cluster spins constitute a small fraction of the total, 
they can accumulate nearly all the doped holes.
The three clearly separated cluster patches in Fig. 3 presumably correspond to one-, two-, and three-dimensional 
clusters, with progressively higher degrees of delocalization and impurity-level band broadening. 

The disproportionate accumulation of holes in impurity-rich cluster regions is simply 
due to preferential filling of cluster impurity states by doped holes. 
Cluster impurity states undergo greater band broadening 
due to stronger effective hopping, and are thus pushed to the valence-band top.
Impurity states corresponding to remaining bulk spins 
lying deeper in the valence band are left with hole deficit. 
With increasing $J$ and more pronounced impurity-level character,
this tendency of differential band broadening is enhanced,
resulting in stronger suppression of ferromagnetism,
highlighting the non-perturbative character of spin couplings in contrast to the 
RKKY picture.\cite{alvarez,dms,diluted,bouzerar}
 
To highlight the macroscopic effects of the disorder induced changes in bulk and cluster spin couplings, 
we compare the magnon density of states (DOS) and temperature dependence of lattice-averaged magnetization 
with the ordered case at the same dilution.
Magnetization was obtained using a locally self-consistent magnon renormalization scheme,\cite{disklm}
which incorporates the spatial character of magnon states as well.
Within the Tyablikov decoupling formalism, similar self-consistent schemes 
involving both global\cite{hilbert} and local\cite{bouz} self-consistency have been employed recently 
in the context of $T_c$ calculations for Heisenberg models with realistic spin couplings.

Fig. 4 (upper panel) shows the characteristic disorder-induced features of 
sharply enhanced low-energy magnon DOS
and high-energy structures due to cluster-localized modes,
associated with the weak (bulk) and strong (cluster) couplings, respectively.
The corresponding features in the magnetization plot (lower panel) are readily identified.
The rapid low-$T$ decay follows from the low-energy magnon softening,
and the subsequent marginally paramagnetic behaviour (more pronounced in some special configurations) 
is due to strong intra-cluster correlations; 
the overall behaviour is characteristic of two distinct (magnon) energy scales.\cite{squid}

In Fig. 4 (lower panel) we have considered a bandwidth $W=12t=10$eV 
of the order of that for GaAs. 
For $J=4t=W/3\approx 3.3$eV as considered in our calculations
and a nominal impurity concentration $x=10\%$, 
the value $Jx = 0.33$eV is close to the corresponding experimental value 
$Jn_{\rm Mn} = J\times 10\% \times n_{\rm Ga} = 
150$meV.\AA$^3$ $\times 10\% \times 2\times 10^{22}$/cm$^3$ $=0.3$eV 
with $J=150$meV.\AA$^3$ taken from Ref. [1]. 
Direct measurements yield $J=1.2\pm 0.2$eV from core-level photoemission\cite{photo} and $J=2.4\pm 0.9$eV from magneto-transport.\cite{matsu,omiya}

In conclusion, we have quantitatively studied the interplay, within a non-perturbative and unified framework, 
of the different ingredients of ferromagnetism in diluted magnets
--- disorder, clustering, fermion polarizations, spin couplings, magnons, and spin dynamics.
Due to the competition between impurities 
for a small fraction of doped holes in the compensated regime $p \ll x$,
impurity disorder plays a highly non-perturbative role.
The direct correlation between the broadly distributed fermion spin polarization values with 
the local measure of spin couplings highlights the essential character of carrier-induced spin couplings
in the presence of disorder, clearly distinguishing between cluster and bulk spins.
Cluster couplings show a large spread for same impurity separation, 
and bulk couplings show distintly non-RKKY character. 

Preferential accumulation of holes in impurity clusters strengthens (weakens) the cluster (bulk) spin couplings,
resulting in softening of low-energy magnons extended over bulk spins
and formation of high-energy cluster-localized modes. 
The locally self-consistent magnon renormalization scheme is able to capture 
these disorder-induced features in the finite temperature spin dynamics.
While the low-energy magnon softening is responsible for the observed 
strong low-$T$ decay of magnetization ($B \approx 1.6\times 10^{-3}$ K$^{-3/2}$ in Fig. 4),
strong intra-cluster correlations tend to prolong global order
through a marginally paramagnetic temperature regime. 
Interestingly, the somewhat slower decay due to cluster correlations 
brings the magnetization fall off closer to the $T^{3/2}$ form over a broader $T$ regime, as observed.\cite{sperl}
Overall, we find that ferromagnetism is considerably weakened compared to the ordered case.
As increasing $p/x$ enhances AF couplings and destabilizes the ferromagnetic state,
reducing spin clustering appears crucial for enhancing ferromagnetism in diluted magnetic semiconductors,
and indeed could be a contributing factor in the annealing process.\cite{ku,sperl}
A two-component picture of bulk and cluster spins, 
with a relatively low carrier density in the bulk could be useful in interpreting 
the relatively high resistivity seen even in the most metallic samples.\cite{matsu,ku,edmonds}

\end{document}